\documentclass[11pt]{article}
\usepackage[preprint]{techreport}   % change to [final] for camera-ready

\title{%
\fontsize{17}{19}\selectfont
DWDP: Distributed Weight Data Parallelism for\\
High-Performance LLM Inference on NVL72
}

\shorttitle{DWDP}   % used in running header
\shortauthor{Author et al.}     % used in running header

\author{%
  Wanqian Li\quad
  Jintao Peng\quad
  Zongfei Jing\quad
  Tianyu Zhang\quad
  Ze Long\quad
  Xianjie Qiao\quad \\
  Xiaoming Chen\quad
  Dongxu Yang\quad
  Kefeng Duan\quad
  June Yang\textsuperscript{1}\\[0.5em]
  NVIDIA
}

\date{\today}

\newcommand{\methodname}{\textsc{DWDP}}  % your method's name

\usepackage{xspace}
\usepackage{xcolor}
\usepackage{float}
\usepackage{subcaption}
\usepackage{listings}
\usepackage{pgfplots}
\pgfplotsset{compat=1.18}

\begin{document}
% -----------------------------------------------------------------------

\maketitle
\renewcommand{\thefootnote}{\arabic{footnote}}
\footnotetext[1]{Corresponding authors:{\texttt{juney@nvidia.com}}}
\setcounter{footnote}{1}

% -----------------------------------------------------------------------
\begin{abstract}
Large language model (LLM) inference increasingly depends on multi-GPU execution,
yet existing inference parallelization strategies require layer-wise inter-rank
synchronization, making end-to-end performance sensitive to workload imbalance.
We present \methodname{} (Distributed Weight Data Parallelism), an inference
parallelization strategy that preserves data-parallel execution while offloading
MoE weights across peer GPUs and fetching missing experts on demand. By removing
collective inter-rank synchronization,
\methodname{} allows each GPU to progress independently. We further address the practical overheads of this
design with two optimizations for split-weight management and asynchronous
remote-weight prefetch.
Implemented in TensorRT-LLM and evaluated with DeepSeek-R1 on GB200 NVL72,
\methodname{} improves end-to-end output TPS/GPU by 8.8\% at comparable TPS/user
in the 20--100 TPS/user serving range under 8K input sequence
length and 1K output sequence length.
\end{abstract}


% =======================================================================
\section{Introduction}
% =======================================================================

Large language models (LLMs) have rapidly advanced across a broad range of
tasks, including natural language understanding, code generation, scientific
reasoning, and multimodal perception~\citep{deepseekr1,mccoy2024embers,gao2025seedance,cheng2025barbarians}. These capability gains have been
accompanied by steady growth in model size~\citep{deepseekv3,kimik25,qwen3coder,glm5}, pushing LLM inference beyond the
memory capacity of a single GPU.
As a result, LLM inference increasingly requires multiple GPUs.

To satisfy the increasing memory demand, LLM inference relies on multiple
forms of model parallelism that partition weights and computation in different
ways, for example, expert parallelism for MoE weights~\citep{deepspeedmoe,tutel,megatron_moe}, tensor parallelism for
hidden dimensions~\citep{megatron}, and pipeline parallelism for
layers~\citep{gpipe}. Although these approaches differ in how they partition
the model, they all
share the same fundamental property: completing each layer requires
cross-GPU coordination. The coordination is implemented through
inter-GPU communication, which introduces synchronization at layer boundaries.

Such layer-wise inter-rank synchronization is increasingly problematic in
real-world LLM serving, where per-rank workloads are rarely balanced. At the
request level, inputs assigned to different ranks often have different
sequence lengths and KV-cache hit rates~\citep{mooncake,strata}, leading to different compute and
memory-access costs. At the weight level, the amount of model computation
activated on each rank can also differ; in MoE models, for example, skewed
routing can make some ranks serve hotter experts and process more tokens than
others. Together, request-level and weight-level imbalance create substantial
per-rank latency variation during inference. Because mainstream parallel
execution strategies synchronize ranks at layer boundaries, local variability
is amplified into global waiting time, and end-to-end throughput becomes
bounded by the slowest rank.

\begin{figure}[H]
  \centering
  \newcommand{\imbfigheight}{1.65in}
  \begin{subfigure}[t]{0.57\linewidth}
    \centering
    \parbox[c][\imbfigheight][c]{\linewidth}{\centering
      \includegraphics[width=\linewidth,height=\imbfigheight,keepaspectratio]{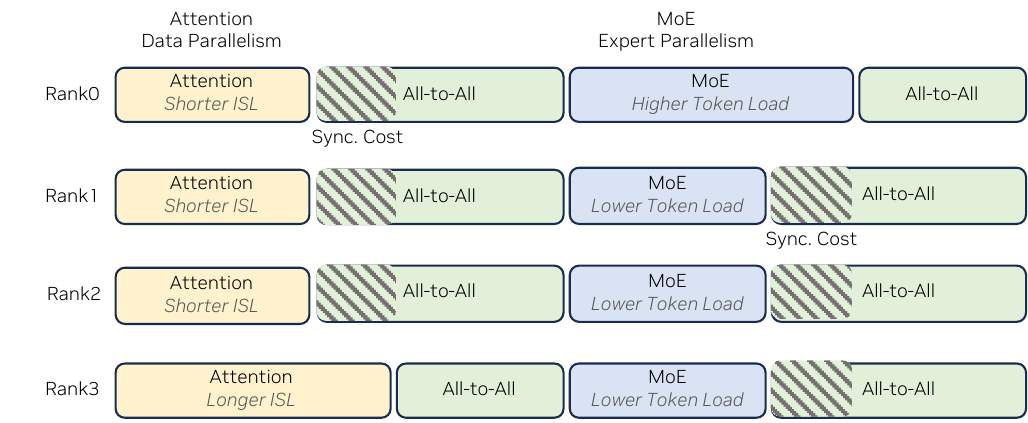}}
    \caption{}
    \label{fig:imbalance-illustration}
  \end{subfigure}
  \hfill
  \begin{subfigure}[t]{0.38\linewidth}
    \centering
    \parbox[c][\imbfigheight][c]{\linewidth}{\centering
      \includegraphics[width=\linewidth,height=\imbfigheight,keepaspectratio]{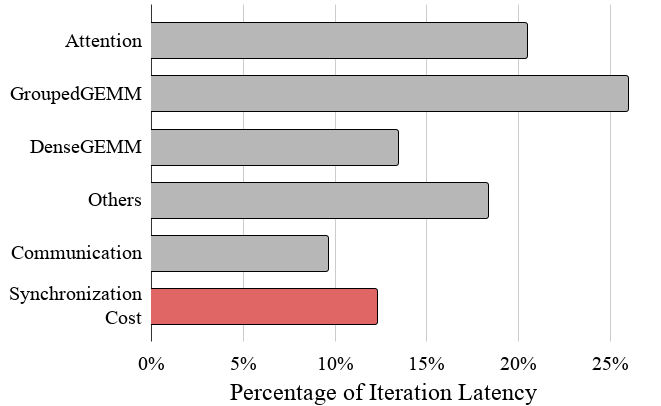}}
    \caption{}
    \label{fig:imbalance-kernel}
  \end{subfigure}
  \caption{Synchronization overhead caused by workload imbalance in DEP. (a)
  Illustration of how request-level and weight-level imbalance are translated
  into waiting time in DEP. (b) Kernel breakdown quantifying the
  synchronization overhead caused by imbalance under DEP. Configuration:
  DeepSeek-R1 on GB200 with input sequence length/output sequence length
  (ISL/OSL) = 8K/1 and input ratio 0.8, meaning that
  the input lengths range from 0.8$\times$8K to 8K.}
  \label{fig:imbalance}
\end{figure}

Figure~\ref{fig:imbalance} illustrates this effect in a configuration that
combines attention data parallelism with expert parallelism, which we denote as
DEP (data parallelism with expert parallelism). Request-level imbalance creates skew after attention, which appears as
waiting time at the first \texttt{all-to-all}. Weight-level imbalance in the
MoE stage then introduces additional skew, which is exposed again at the
second \texttt{all-to-all}. The kernel breakdown in
Figure~\ref{fig:imbalance} shows that
synchronization overhead reaches approximately 12\% when the coefficient of
variation of per-rank sequence lengths is 20\%---a level of imbalance well
within the range observed in production workloads. These results show that
synchronization overhead is not a corner case: under realistic imbalance, it
materially reduces end-to-end inference throughput.

Prior work reduces inter-rank imbalance through better scheduling and load
balancing, such as cache-aware scheduling~\citep{preble,sglang_v04,mooncake,strata,tensorrtllm_adp}, load-aware scheduling~\citep{distserve,splitwise}, and expert
load balancing~\citep{tutel,tensorrtllm_ep}. However, these techniques do not fundamentally change the
synchronization requirement of existing inference parallelization strategies. As
serving workloads become more diverse, this synchronization overhead remains
unavoidable and continues to limit performance.

To address this limitation, we present \methodname{} (Distributed Weight
Data Parallelism), an inference parallelization strategy that removes
collective inter-rank synchronization from the critical inference path. In
\methodname{}, ranks remain data-parallel, while MoE weights are offloaded
across peer GPUs to collectively accommodate the full model and are fetched on
demand during inference. As a result, each rank can execute inference
independently without waiting for other ranks, which directly mitigates the
performance loss caused by workload imbalance. DWDP is most effective when
remote-weight prefetch can be hidden behind computation, which in practice
requires both high-bandwidth peer-to-peer GPU communication and sufficiently
large layer-wise compute windows. These conditions are often met in the
context phase of LLM inference, motivating our focus on applying
DWDP to the context servers.

\paragraph{Contributions.}
\begin{enumerate}[leftmargin=1.5em]
  \item We present \methodname{}, an inference parallelization strategy that
  combines data parallelism with GPU-based weight offloading. \methodname{}
  removes collective inter-rank synchronization from the critical inference
  path, allowing each rank to execute inference independently without waiting
  for other ranks.

  \item We identify two key challenges in realizing \methodname{} efficiently:
  split-weight management overhead caused by offloading, and communication-side
  performance loss during asynchronous remote-weight prefetch. We propose two
  corresponding optimizations and analyze the residual interference caused by
  communication-computation overlap.

  \item We implement \methodname{} in TensorRT-LLM and evaluate it with
  DeepSeek-R1 on GB200 NVL72. In typical 8K/1K serving scenarios within the
  20--100 TPS/user range, \methodname{} improves end-to-end output TPS/GPU by
  8.8\% at comparable TPS/user and pushes the performance frontier to a
  better region.
\end{enumerate}

\section{Methodology Overview}

\begin{figure}[t]
  \centering
  \includegraphics[width=0.98\linewidth]{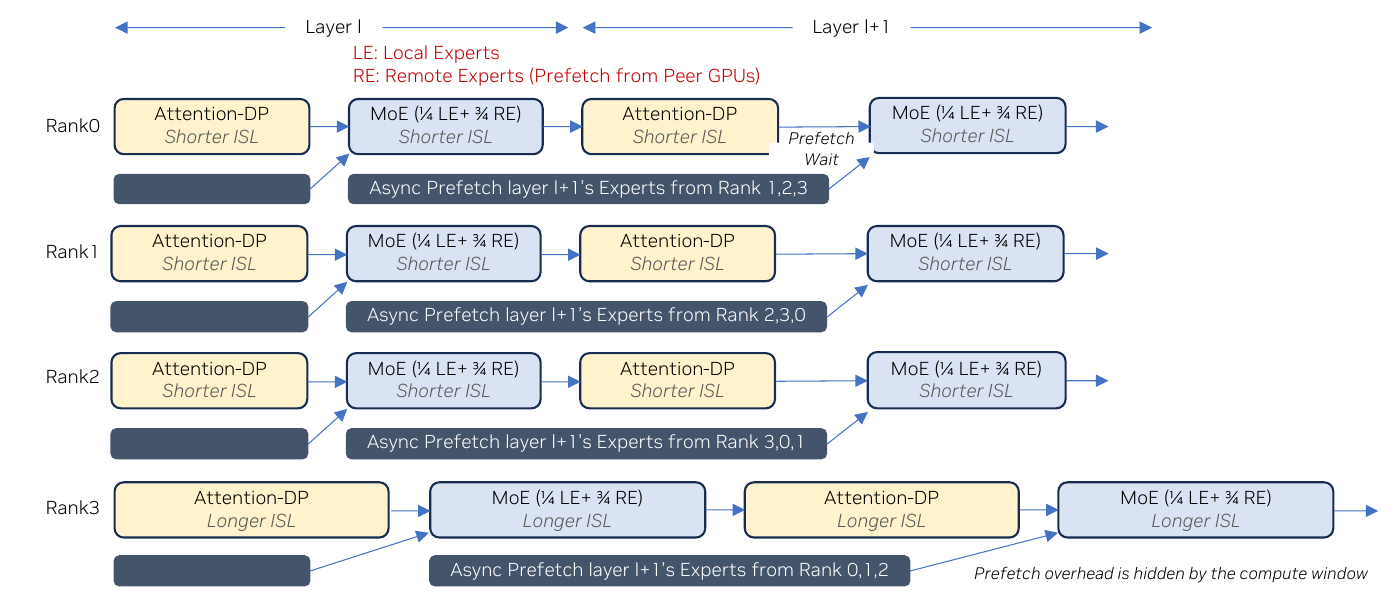}
  \vspace{0.5em}
  \caption{Overview of \methodname{} with DWDP group size 4.}
  \label{fig:dwdp-overview}
\end{figure}

Figure~\ref{fig:dwdp-overview} illustrates the core idea of \methodname{} on an
MoE-based LLM~\citep{deepseekv3,deepseekr1}.
To accommodate the full model, \methodname{} preserves data-parallel execution
across ranks while offloading MoE weights across peer GPUs. This design targets
MoE weights because they dominate the model memory footprint, whereas attention
weights account for only a relatively small fraction.

Within a DWDP group, attention weights are fully replicated on each rank,
whereas the experts in every MoE layer are partitioned across ranks. As a
result, each rank permanently stores only its local experts, while the remaining
experts reside on peer GPUs. Before executing an MoE layer, the rank fetches the
missing remote experts on demand to assemble the weights of one layer.

At runtime, \methodname{} overlaps the asynchronous prefetch of the missing
remote experts for layer $l+1$ with the MoE block of layer $l$ and the
attention block of layer $l+1$. Together, these two blocks provide the compute
window for hiding remote weight prefetch.
Before executing the MoE block of layer $l+1$, the rank waits until all remote
experts have arrived. After the layer finishes, the prefetched remote experts
are released. To sustain this pipeline across layers, \methodname{} uses double
buffering for prefetching.

To eliminate inter-rank synchronization during inference,
\methodname{} avoids NCCL-based collectives~\citep{nccl}, such as all-gather, for remote
weight gathering, as these operations synchronize ranks.
Instead, each rank pulls remote experts from peer GPUs via copy-engine-based
\texttt{cudaMemcpyAsync}, which does not occupy SM resources. The transfers are
issued as serial peer-to-peer pulls, which avoid introducing synchronization
across ranks. As a result, each rank can execute inference independently
without waiting for other ranks once the required experts for a layer have
arrived. From the perspective of disaggregated serving~\citep{distserve,splitwise,mooncake}, each rank remains an
independent inference worker that can receive requests and return responses
independently.

Beyond enabling fully asynchronous inference, DWDP also provides greater
flexibility in expert placement. Because each rank only needs to fetch the
weights of one layer before executing its MoE block, DWDP does not require the
number of experts to be exactly divisible by the DWDP group size, nor does it
require a perfectly disjoint expert partition across ranks. Instead, ranks may
be configured with the same number of local experts while allowing redundant
expert placement when necessary, for example, to support group sizes that do
not evenly divide the number of experts. This weak placement constraint enables
resource provisioning at single-rank granularity. When memory permits, the same
redundancy can also reduce remote prefetch overhead by increasing the number of
local experts on each rank.

Together, fully asynchronous inference and flexible expert placement are the
two main benefits of DWDP. However, this new strategy also introduces new
challenges. Because weights are split into local experts and remote experts,
DWDP requires efficient split-weight management. In addition, asynchronous
remote-weight prefetch can introduce communication-side overheads, including
communication-computation interference and many-to-one contention at the source
rank. Section~\ref{sec:design-optimizations} will present the corresponding optimizations in detail.

\section{Preliminary Analysis}

We use a layer-wise roofline model~\citep{williams2009roofline} to identify when \methodname{} can
outperform DEP and what fundamentally limits its gain. In this analysis, we focus on the context
phase of DeepSeek-R1 on GB200 and compare DWDP4 against DEP4\footnote{Here, the
suffix 4 denotes a four-rank execution group. DWDP4 uses \methodname{} with
DWDP group size 4, while DEP4 uses the DEP baseline with the same four-rank
group size.}.

At a high level, DWDP can overlap remote expert prefetch with computation,
whereas DEP performs computation and expert-parallel communication
synchronously. We therefore model the per-layer latency of DWDP as
$T_{\mathrm{DWDP}}=\max(T_{\mathrm{compute}}, T_{\mathrm{prefetch}})$ and that
of DEP as $T_{\mathrm{DEP}}=T_{\mathrm{compute}}+T_{\mathrm{all2all}}$, where
$T_{\mathrm{prefetch}}$ denotes the time to fetch missing remote expert weights
and $T_{\mathrm{all2all}}$ denotes the expert-parallel \mbox{all-to-all}
latency.

To estimate $T_{\mathrm{compute}}$, we first model each operator in a layer
using a standard roofline approximation. For an operator with FLOPs $F$ and
memory traffic $B$, we estimate its latency as
$T_{\mathrm{op}}=\max(F/P_{\mathrm{peak}}, B/\mathrm{BW}_{\mathrm{mem}})$,
where $P_{\mathrm{peak}}$ and $\mathrm{BW}_{\mathrm{mem}}$ are the hardware
compute throughput and memory bandwidth, respectively. Summing the operator
times for attention and MoE yields $T_{\mathrm{compute}}$.

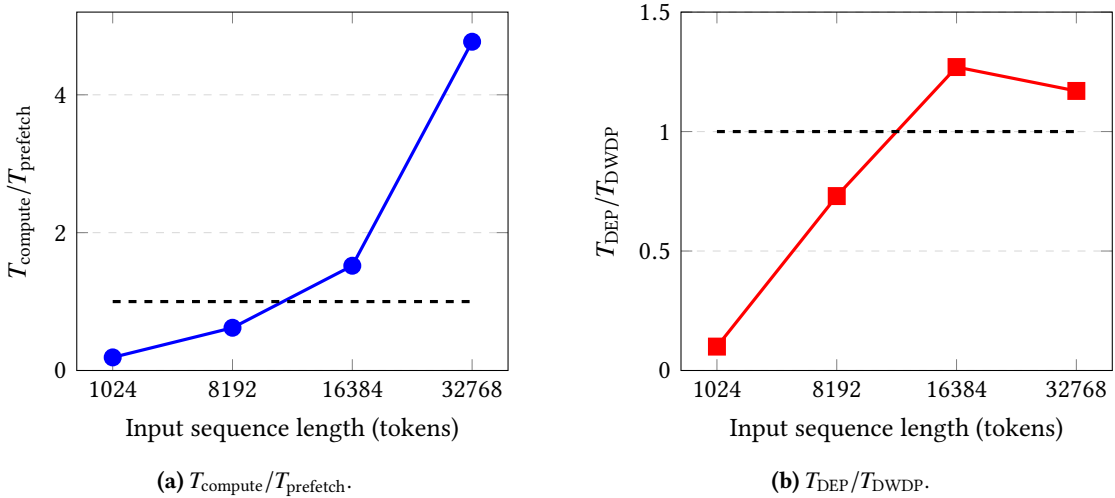
\begin{figure}[H]
  \centering
  \begin{subfigure}[t]{0.48\linewidth}
    \centering
    \begin{tikzpicture}
      \begin{axis}[
        width=\linewidth,
        height=2.5in,
        xlabel={Input sequence length (tokens)},
        ylabel={$T_{\mathrm{compute}}/T_{\mathrm{prefetch}}$},
        ymin=0,
        ymax=5.2,
        symbolic x coords={1024,8192,16384,32768},
        xtick=data,
        ymajorgrids=true,
        grid style={dashed,gray!30},
        tick label style={font=\small},
        label style={font=\small},
        every axis plot/.append style={line width=1.2pt, mark size=2.8pt},
      ]
        \addplot[color=blue, mark=*] coordinates {
          (1024,0.19)
          (8192,0.62)
          (16384,1.52)
          (32768,4.77)
        };
        \addplot[color=black, dashed, forget plot] coordinates {
          (1024,1)
          (32768,1)
        };
      \end{axis}
    \end{tikzpicture}
    \caption{$T_{\mathrm{compute}}/T_{\mathrm{prefetch}}$.}
  \end{subfigure}
  \hfill
  \begin{subfigure}[t]{0.48\linewidth}
    \centering
    \begin{tikzpicture}
      \begin{axis}[
        width=\linewidth,
        height=2.5in,
        xlabel={Input sequence length (tokens)},
        ylabel={$T_{\mathrm{DEP}}/T_{\mathrm{DWDP}}$},
        ymin=0,
        ymax=1.5,
        symbolic x coords={1024,8192,16384,32768},
        xtick=data,
        ymajorgrids=true,
        grid style={dashed,gray!30},
        tick label style={font=\small},
        label style={font=\small},
        every axis plot/.append style={line width=1.2pt, mark size=2.8pt},
      ]
        \addplot[color=red, mark=square*] coordinates {
          (1024,0.10)
          (8192,0.73)
          (16384,1.27)
          (32768,1.17)
        };
        \addplot[color=black, dashed, forget plot] coordinates {
          (1024,1)
          (32768,1)
        };
      \end{axis}
    \end{tikzpicture}
    \caption{$T_{\mathrm{DEP}}/T_{\mathrm{DWDP}}$.}
  \end{subfigure}
  \caption{Roofline-based preliminary analysis for the DeepSeek-R1 context phase on GB200,
  comparing DWDP4 against DEP4 at batch size 1. The two subplots separately
  show the compute-to-prefetch ratio and the DEP-to-DWDP runtime ratio. The
  dashed line at $y=1$ marks the boundary where prefetch can be fully hidden
  and where DWDP begins to outperform DEP.}
  \label{fig:roofline-adv-limit}
\end{figure}

Based on this model, we focus on two derived metrics: the ratio
$T_{\mathrm{compute}}/T_{\mathrm{prefetch}}$, which indicates whether DWDP can
hide remote weight prefetch, and the speedup $T_{\mathrm{DEP}}/T_{\mathrm{DWDP}}$,
which captures DWDP's expected advantage over DEP.
Figure~\ref{fig:roofline-adv-limit} plots both metrics across input sequence
lengths (ISLs) at batch size 1.

We observe that DWDP begins to outperform DEP at around 16K tokens. As ISL
increases, $T_{\mathrm{compute}}/T_{\mathrm{prefetch}}$ grows from below 1
to above 1, indicating that longer contexts provide a sufficiently large
compute window to amortize and eventually hide remote prefetch overhead. This
reveals that DWDP requires enough computation per layer to
cover remote weight prefetch.
This 16K threshold is specific to the batch-size-1 setting. Increasing the batch
size enlarges the compute window and can make DWDP beneficial even for shorter
contexts.

DWDP's advantage over DEP comes from eliminating synchronized
all-to-all communication from the critical path. This
advantage, however, is not monotonic in ISL. Once the sequence becomes very
long, computation dominates both methods, so the synchronized
\mbox{all-to-all} overhead accounts for a smaller fraction of DEP's latency.
Accordingly, the marginal speedup of DWDP decreases as ISL grows further.

This analysis assumes perfectly balanced workloads. On the one hand, it does not
capture the additional benefit of DWDP in reducing synchronization overhead under imbalanced workloads. On the other hand, it also
omits practical issues in implementing DWDP, such as contention when weight prefetch overlaps with computation. We will discuss these issues and the
corresponding optimizations in the next section.

\section{Design Challenges and Optimizations}
\label{sec:design-optimizations}

This section presents the key implementation overheads of \methodname{} and the
techniques we use to address them. In particular, we eliminate split-weight
merge overhead and mitigate many-to-one source-side communication contention
caused by asynchronous prefetch. We first conduct the kernel breakdown and
runtime-trace analysis of a baseline implementation, and then introduce the
corresponding optimizations guided by this analysis.

\subsection{Baseline Implementation and Profiling}
\label{sec:baseline-profiling}

\subsubsection{Kernel Breakdown Analysis}

To understand the practical bottlenecks of \methodname{}, we first evaluate a naive baseline implementation. Table~\ref{tab:context-breakdown} compares the context-only iteration-latency breakdown of DEP4 and DWDP4 for DeepSeek-R1.

\begin{table}[H]
  \centering
  \small
  \caption{Context-only iteration-latency breakdown of DEP4 and DWDP4 for DeepSeek-R1 under input sequence length$=8$K, ratio$=0.8$, and context-phase maximum number of tokens$=32768$ on GB200. The last column reports per-category deltas normalized to the DEP4 iteration latency.}
  \label{tab:context-breakdown}
  \begin{tabular}{lrrr}
    \toprule
    Category & DEP4 ($\mu$s) & DWDP4 ($\mu$s) & $\Delta / T_{\mathrm{DEP4}}$ \\
    \midrule
    Attention & 269.67 & 320.56 & -3.86\% \\
    GroupedGEMM & 342.40 & 337.42 & 0.38\% \\
    DenseGEMM & 177.50 & 189.28 & -0.89\% \\
    Others & 241.69 & 284.32 & -3.23\% \\
    Communication & 126.74 & 0.00 & 9.60\% \\
    D2D Copy & 0.00 & 34.00 & -2.58\% \\
    P2P Copy & 0.00 & 429.00 & -- \\
    Synchronization Cost & 161.85 & 0.00 & 12.26\% \\
    \midrule
    \textbf{Iteration Latency} & \textbf{1319.85} & \textbf{1165.58} & \textbf{11.69\%} \\
    \bottomrule
  \end{tabular}
\end{table}

The breakdown highlights both the promise and the remaining inefficiencies of \methodname{}. Relative to DEP, DWDP removes synchronization cost entirely and takes communication off the critical path. Together, these two effects correspond to a 21.86\% gross reduction in iteration latency. 
At the same time, the baseline implementation exhibits two regressions. First, it incurs D2D-copy overhead. Second, compute categories such as Attention and Others become slower. Together, these regressions reduce the realized gain to a net 11.69\% improvement.

This breakdown motivates the rest of this section. First, the baseline DWDP implementation introduces $34.00\,\mu$s of D2D copy time because local and prefetched remote weights must be merged into a contiguous buffer before computation; Section~\ref{sec:eliminate-merge} addresses this overhead. Second, the increased Attention and Others times show that overlapping remote-weight prefetch with computation introduces non-negligible communication-computation interference. We analyze this effect in Appendix~\ref{sec:appendix-interference} and show that, in our setting, the dominant cause is power-induced frequency throttling.

\subsubsection{Runtime-Trace Analysis}
\label{sec:runtime-trace-analysis}

Beyond the kernel breakdown, we further analyze Nsight Systems runtime traces of the baseline DWDP implementation. This analysis reveals a practical issue of asynchronous remote-weight pulls: in each MoE layer, different destination ranks can simultaneously pull missing remote experts from the same source rank, creating many-to-one communication contention at the source-side copy engine.

\begin{figure}[H]
  \centering
  \includegraphics[width=\linewidth]{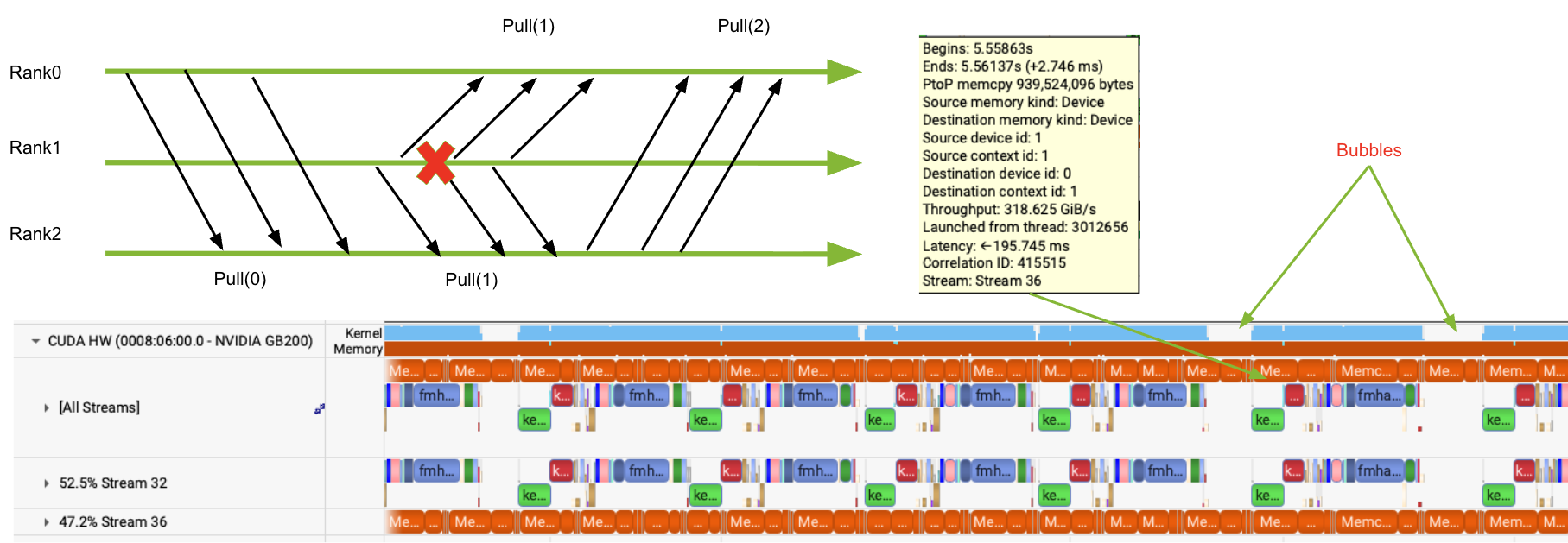}
  \caption{Nsight Systems trace showing many-to-one source-side communication contention in DWDP under max\_num\_tokens$=16384$ and input sequence lengths ranging from 4K to 8K. Multiple destination ranks concurrently pull remote weights from the same source rank, so the source-side copy engine serializes these requests and exposes compute bubbles.}
  \label{fig:async-communication-contention}
\end{figure}

Figure~\ref{fig:async-communication-contention} shows a representative case in which the layer-wise compute window is only comparable to the remote-weight prefetch time. In this regime, serialization at the source rank stretches the communication window and exposes visible compute bubbles before the next compute region can begin. Although this effect does not appear as a separate category in Table~\ref{tab:context-breakdown}, it can still materially reduce the realized gain of DWDP. Section~\ref{sec:solve-contention} addresses this source-side serialization with a time-division multiplexing scheme.

\subsection{Eliminating Split-Weight Merge Overhead}
\label{sec:eliminate-merge}

DWDP naturally produces split weights for each MoE layer: each rank
permanently stores its local experts and fetches missing remote experts into
separate buffers on demand. As a result, the weights required
by the MoE computation are not stored in a single contiguous buffer.

This layout is problematic for existing MoE groupedGEMM kernels, which
typically assume a single contiguous weight buffer. In a straightforward
implementation, the runtime must first merge local and prefetched remote
weights through a pre-launch device-to-device copy before launching the kernel. This extra
step increases memory-bandwidth consumption and latency overhead.

To eliminate this merge overhead, we extend the MoE groupedGEMM kernel to
consume multiple weight buffers directly.\footnote{Our groupedGEMM kernel is
implemented in CuTeDSL~\citep{cute_dsl}. Its extensible interface makes this change
straightforward.} Specifically, we add TensorList-based inputs so that the
kernel can select from local and prefetched remote buffers internally, rather
than relying on an externally merged buffer. This design remains compatible
with existing layouts and sharding schemes. Although it introduces modest
additional indexing and address-computation overhead, profiling and end-to-end
evaluation show no meaningful performance regression.

\subsection{Mitigating Asynchronous Communication Contention}
\label{sec:solve-contention}

\subsubsection{Theoretical Analysis}
The runtime traces in Section~\ref{sec:runtime-trace-analysis} show that fully asynchronous remote-weight pulls can create many-to-one contention at the source rank and expose compute bubbles. To isolate the effect of this contention, we analyze the boundary regime in which the ideal communication time is comparable to the layer-wise compute time. In this regime, communication would be nearly hidden without contention, so the key question is whether asynchronous contention alone can turn nearly hidden communication into exposed compute bubbles.

Consider a DWDP group with $N$ ranks. In one communication round, each rank must pull remote weights from the other $N-1$ ranks. Let the ideal total communication time without contention be $T$. Since each rank issues $N-1$ serial pulls, the ideal service time of one pull is
\begin{equation}
\tau = \frac{T}{N-1}.
\end{equation}

We adopt a random-state model: whenever a rank becomes ready to issue its next pull, its source rank is uniformly distributed over the remaining peers. For a tagged rank, after it selects one source, each of the other $N-2$ ranks chooses the same source with probability $1/(N-1)$. Therefore, the number of competing requests for that source is
\begin{equation}
X \sim \mathrm{Binomial}\left(N-2, \frac{1}{N-1}\right).
\end{equation}
To simplify the model, we only compute the probability of contention. Define the contention level $C$ as the total number of pulls that concurrently target the same source, including the tagged pull itself. Then $C = X + 1$.

We stop at the contention probability rather than translating it into exact communication time. Under a fully serialized equal-size approximation, a pull with contention degree $C$ would incur roughly $C\tau$ latency, but the actual latency depends on many system factors. Our goal here is only to show that low-order many-to-one contention arises naturally and therefore should be explicitly mitigated.
Table~\ref{tab:contention-probabilities} summarizes the contention probabilities for several DWDP group sizes.
\begin{table}[H]
  \centering
  \caption{Contention probability $\Pr[C=c]$ under the random asynchronous model. Entries are percentages.}
  \label{tab:contention-probabilities}
  \resizebox{\linewidth}{!}{
  \begin{tabular}{lccccccccccccccc}
    \toprule
    Config & $C=1$ & $C=2$ & $C=3$ & $C=4$ & $C=5$ & $C=6$ & $C=7$ & $C=8$ & $C=9$ & $C=10$ & $C=11$ & $C=12$ & $C=13$ & $C=14$ & $C=15$ \\
    \midrule
    DWDP3  & 50.00 & 50.00 & -- & -- & -- & -- & -- & -- & -- & -- & -- & -- & -- & -- & -- \\
    DWDP4  & 44.44 & 44.44 & 11.11 & -- & -- & -- & -- & -- & -- & -- & -- & -- & -- & -- & -- \\
    DWDP6  & 40.96 & 40.96 & 15.36 & 2.56 & 0.16 & -- & -- & -- & -- & -- & -- & -- & -- & -- & -- \\
    DWDP8  & 39.66 & 39.66 & 16.52 & 3.67 & 0.46 & 0.03 & 0.00085 & -- & -- & -- & -- & -- & -- & -- & -- \\
    DWDP12 & 38.55 & 38.55 & 17.35 & 4.63 & 0.81 & 0.097 & 0.0081 & 0.00046 & 0.000017 & $3.86\times10^{-7}$ & $3.86\times10^{-9}$ & -- & -- & -- & -- \\
    DWDP16 & 38.06 & 38.06 & 17.67 & 5.05 & 0.99 & 0.14 & 0.015 & 0.0012 & 0.000077 & $3.69\times10^{-6}$ & $1.32\times10^{-7}$ & $3.42\times10^{-9}$ & $6.11\times10^{-11}$ & $6.71\times10^{-13}$ & $3.43\times10^{-15}$ \\
    \bottomrule
  \end{tabular}}
\end{table}

These results show that under random asynchronous execution, the most likely cases are low-order contentions such as $C=1$ and $C=2$, but the probability mass of higher-order contentions grows gradually with $N$. Therefore, even without a pathological schedule, larger DWDP groups naturally face a higher chance of many-to-one communication contention. This observation motivates a mitigation that improves robustness precisely in the common low-order regime.

\subsubsection{Copy with Time-Division Multiplexing}
To solve many-to-one contention at the source rank, we split each remote-weight transfer into fixed-size slices and schedule these slices in a round-robin manner across all active destination ranks.
% 是否会暴露NV底层硬件细节
The key architectural observation is that the copy engine is pipelined and, for sufficiently small requests, it can keep small slices in flight.
Let the slice size be $s$ and consider the regime where $s$ is small enough that the copy engine can keep two slices in flight at the same time. Under this assumption, once one slice has been issued, the engine can begin issuing another small slice before the first one fully retires. Therefore, the source can continue making progress on another request even if one slice is slowed down by communication contention. In other words, chunking converts a monolithic pull into a sequence of smaller DMA requests, and pipelined execution enables partial overlap among them.

Our implementation maintains one pending-slice queue for each destination rank at every source rank. When a rank needs to pull a remote expert shard of size $M$, the shard is partitioned into $K = \left\lceil \frac{M}{s} \right\rceil$ slices.
The source-side scheduler serves all non-empty queues in round-robin order and issues at most one slice from each queue in one round.
Listing~\ref{lst:batched-prefetch-copy-plan} shows the pseudocode for building such a batched prefetch-copy plan. The key idea is to iterate over slice offsets first and then traverse peers in round-robin order, so that slices from different source ranks are interleaved in the final DMA schedule.
\begin{lstlisting}[language=Python,basicstyle=\ttfamily\footnotesize,columns=fullflexible,keepspaces=true,showstringspaces=false,breaklines=true,frame=single,caption={Pseudocode for building a batched prefetch-copy plan in round-robin slice order.},label={lst:batched-prefetch-copy-plan}]
copy_plan = []
for each parameter p:
    M = prefetch_size(p)
    for offset = 0 to M step s:
        chunk = min(s, M - offset)
        for peer in round_robin(remote_peers):
            src = peer_ptr(peer, p) + src_offset(peer, p) + offset
            dst = local_buffer(peer, p) + offset
            copy_plan.append((dst, src, chunk))
return copy_plan
\end{lstlisting}
Conceptually, this mechanism uses the pipelined execution of small requests to create finer-grained destination-side traffic partitioning. Instead of letting one destination monopolize the copy engine with a large monolithic pull, slicing and round-robin force the source to interleave progress across destinations at slice granularity. This effect is similar in spirit to virtual-lane-based isolation in high-performance networks: when one lane is slowed down by contention, the remaining lanes can still continue utilizing the network at high efficiency. Here, the per-destination slice queues play a similar role, preventing one blocked pull from stalling the progress of all other pulls targeting the same source.

Under this setting, a particularly important consequence follows. If the pipelined engine can keep two small slices in flight at the same time, then the rank-level data pull does not slow down even when both in-flight requests each see contention degree $2$. Equivalently, communication-induced slowdown appears only when both in-flight slices simultaneously encounter contention degree at least $3$. This provides an intuitive explanation for why two-slice pipelining is robust against low-order contention: one mildly contended slice can still keep the port busy while the other slice is temporarily delayed. At the same time, because network fluctuation is unavoidable in practice, contention on small slices is typically short-lived and is less likely to block the network for an extended period of time.

\section{Experimental Results}
\label{sec:experimental-results}
\subsection{Experimental Setup}

\paragraph{Hardware, Software Stack, and Model}
All experiments are conducted on GB200 NVL72~\citep{nvidia_gb200} platform. We implement \methodname{} in TensorRT-LLM~\citep{tensorrtllm_repo} inference
framework\footnote{Based on TensorRT-LLM commit 3a89495. The upstream
integration is in progress via \href{https://github.com/NVIDIA/TensorRT-LLM/pull/12136/changes}{TensorRT-LLM PR \#12136}.}. We
evaluate DeepSeek-R1~\citep{deepseekr1,deepseekv3} using the NVFP4 checkpoint, where MoE weights are quantized
in NVFP4 precision and the attention module uses FP8 KV cache.

\paragraph{Baselines}
All evaluations are conducted in the disaggregated serving mode~\citep{distserve,splitwise,mooncake}, where
\methodname{} is applied to the context server. Our primary baseline is a
conventional DEP-based configuration under the same runtime and
hardware constraints. Unless otherwise stated, we keep the generation-server
configuration unchanged and only modify the context-server configuration related to \methodname{}.

\paragraph{Metrics}
For context-only experiments, we primarily report TPS/GPU speedup and time to
first token (TTFT) speedup over the DEP baseline. For end-to-end experiments,
we report tokens per second per user (TPS/user), tokens per second per GPU
(TPS/GPU), and TTFT. In particular, we focus on the throughput-efficiency tradeoff
at particular TPS/user constraints. TTFT is reported as the median time from
request arrival to the first generated token, including queueing time. Together,
these metrics let us characterize both the performance gains of \methodname{}
in the context phase and their end-to-end impact on serving efficiency.

\subsection{Context-Only Results}
\paragraph{Context-Only Setup}
This subsection evaluates the isolated context phase using the Artificial
Analysis dataset. We first study how the performance gain of \methodname{}
varies with workload characteristics and server configurations. We then
isolate the contribution of the two optimizations introduced in
Section~\ref{sec:design-optimizations}.
Together, these studies validate whether the asynchronous execution model of
\methodname{} improves context-server efficiency under different workload
characteristics and server configurations.
Unless otherwise stated, each experiment changes only one factor at a time,
while the remaining parameters are fixed and listed in the corresponding
table.

\paragraph{Ablation Study on Workload and Configuration}
We first study how the context-only performance speedup of \methodname{}
varies with ISL, the context-phase maximum number of tokens (MNT), workload
imbalance, and DWDP group size.
Unless otherwise stated, the results in this paragraph do not yet include the
contention-mitigation optimization in
Section~\ref{sec:design-optimizations}; its contribution is evaluated
separately later in this subsection.
\begin{table}[t]
  \centering
  \caption{Ablation study of context-only performance under different workload
  and configuration settings.}
  \label{tab:context-ablation}

  \begin{subtable}[t]{0.48\linewidth}
    \centering
    \caption{Performance speedup vs. ISL (fixed MNT=32768).}
    \label{tab:context-speedup-isl}
    \footnotesize
    \begin{tabular}{lcc}
      \toprule
      ISL & TTFT speedup & TPS/GPU speedup \\
      \midrule
      1024 & 1.27 & 1.11 \\
      8192 & 1.16 & 1.10 \\
      16384 & 1.12 & 1.09 \\
      32768 & 1.11 & 1.09 \\
      \bottomrule
    \end{tabular}
  \end{subtable}
  \hfill
  \begin{subtable}[t]{0.48\linewidth}
    \centering
    \caption{Performance speedup vs. MNT (fixed ISL=8192).}
    \label{tab:context-speedup-max-tokens}
    \footnotesize
    \begin{tabular}{lcc}
      \toprule
      MNT & TTFT speedup & TPS/GPU speedup \\
      \midrule
      16384 & 1.07 & 1.01 \\
      32768 & 1.16 & 1.10 \\
      \bottomrule
    \end{tabular}
  \end{subtable}

  \vspace{0.8em}

  \begin{subtable}[t]{0.48\linewidth}
    \centering
    \caption{Performance speedup vs. ISL/STD.}
    \label{tab:context-speedup-isl-std}
    \footnotesize
    \begin{tabular}{lcc}
      \toprule
      ISL/STD & TTFT speedup & TPS/GPU speedup \\
      \midrule
      16384/0 & 1.12 & 1.09 \\
      16384/1024 & 1.11 & 1.08 \\
      16384/2048 & 1.13 & 1.11 \\
      16384/4096 & 1.18 & 1.15 \\
      \bottomrule
    \end{tabular}
  \end{subtable}
  \hfill
  \begin{subtable}[t]{0.48\linewidth}
    \centering
    \caption{Performance speedup vs. DWDP group size \\(fixed ISL=16384, MNT=32768).}
    \label{tab:context-speedup-dwdp-size}
    \footnotesize
    \begin{tabular}{lcc}
      \toprule
      Group size & TTFT speedup & TPS/GPU speedup \\
      \midrule
      DWDP3 & 0.86 & 1.093 \\
      DWDP4 & 1.15 & 1.091 \\
      \bottomrule
    \end{tabular}
  \end{subtable}
\end{table}

Table~\ref{tab:context-speedup-isl} shows that \methodname{} achieves
1.09--1.11$\times$ TPS/GPU speedup and 1.11--1.27$\times$ TTFT speedup
across ISLs from 1K to 32K. When MNT is fixed and sufficiently large, the
performance speedup decreases as ISL increases. This trend is consistent with
the preliminary analysis in Figure~\ref{fig:roofline-adv-limit}: as the
sequence length grows, computation accounts for a larger fraction of the
context-phase latency, so the relative gain from reducing synchronized
communication becomes smaller. At the same time, these results do not contradict
the batch-size-1 threshold identified in the preliminary analysis. Here MNT is fixed at 32768, so the
runtime can form larger effective context-phase batches; this enlarges the
compute window and allows even 1K inputs to benefit from DWDP.

Table~\ref{tab:context-speedup-max-tokens} shows that \methodname{} achieves
1.01--1.10$\times$ TPS/GPU speedup and 1.07--1.16$\times$ TTFT speedup
across different MNT settings. With the same ISL, a larger runtime token
budget per forward pass leads to higher performance speedup because it creates
a larger compute window for hiding weight prefetch overhead.

Table~\ref{tab:context-speedup-isl-std} shows that \methodname{} achieves
1.08--1.15$\times$ TPS/GPU speedup and 1.11--1.18$\times$ TTFT speedup as
workload imbalance increases. As the standard deviation of input sequence
lengths grows, \methodname{} achieves higher speedup over DEP. This result
supports the core motivation of DWDP: under more imbalanced workloads, the
asynchronous execution of \methodname{} avoids the growing synchronization cost
incurred by DEP.

Table~\ref{tab:context-speedup-dwdp-size} shows that DWDP3 and DWDP4 deliver
nearly identical TPS/GPU speedup, indicating that the core context-side
benefit of \methodname{} is preserved at smaller group sizes. DWDP3, however,
shows worse TTFT speedup, likely because the smaller context-side deployment
provides lower aggregate throughput and therefore higher queueing delay before
the first token. More importantly, DWDP3 highlights an additional advantage of
\methodname{}: it supports finer-grained resource allocation than conventional
DEP configurations, which can be useful in disaggregated serving.

Overall, these ablations show that \methodname{} consistently improves
context-side performance. Its gains are largest when the workload provides a
sufficiently large compute window to hide weight prefetch overhead and when
workload imbalance makes DEP's synchronization cost more severe.

\paragraph{Evaluation of Split-Weight Merge Elimination}
We evaluate the split-weight merge elimination optimization with the same configuration as Table~\ref{tab:context-breakdown}.
With this optimization, \methodname{} improves TPS/GPU by about 3\% over the DWDP baseline by removing the D2D merge copy from the critical path, without introducing a meaningful regression in groupedGEMM execution.

\paragraph{Evaluation of Contention Mitigation}
We evaluate contention mitigation under the ISL=8K context-only workload using 1MB slices, varying ISL ratio and MNT. Table~\ref{tab:tdm-e2e-throughput} compares context-only TPS/GPU normalized to DEP between DWDP with split-weight merge elimination only and full \methodname{} with contention mitigation.
The additional gain from contention mitigation is largest when the layer-wise compute window is short, i.e., when MNT is smaller and ISL ratio is lower.

\begin{table}[t]
  \centering
  \caption{Context-only TPS/GPU under the ISL=8K workload using 1MB slices, normalized to DEP within each row. The last two columns compare DWDP with and without contention mitigation after split-weight merge elimination.}
  \label{tab:tdm-e2e-throughput}
  \resizebox{0.72\linewidth}{!}{
  \begin{tabular}{lccccc}
    \toprule
    ISL Ratio & MNT & DEP & DWDP + Merge Elim. & Full \methodname{} \\
    \midrule
    0.5 & 16384 & 1.000 & 0.995 & 1.081 \\
    0.5 & 32768 & 1.000 & 1.140 & 1.139 \\
    0.8 & 16384 & 1.000 & 1.039 & 1.053 \\
    0.8 & 32768 & 1.000 & 1.098 & 1.109 \\
    \bottomrule
  \end{tabular}}
\end{table}

Table~\ref{tab:tdm-e2e-throughput} shows that the main performance gain of contention mitigation comes from eliminating the random communication delays that remain after split-weight merge elimination. This effect is strongest when the context-side compute window is short. Accordingly, the benefit of the full design is larger when MNT is smaller and when the ISL ratio is lower, because a smaller average ISL shortens the layer-wise compute time while leaving the communication time largely unchanged. This trend is clearest at ISL ratio $=0.5$ and MNT $=16$K, where DWDP with merge elimination is slightly below DEP while full \methodname{} is substantially higher. As the compute window becomes larger, for example at MNT $=32$K, the additional benefit becomes much smaller because the baseline configuration can already hide a larger fraction of the communication cost.

\subsection{End-to-End Results}
\paragraph{End-to-End Setup}

This subsection evaluates the end-to-end performance impact of
\methodname{} under
disaggregated serving. We use the SemiAnalysis dataset with maximum input
length 8K, output length 1K, and input lengths ranging from 6.4K to 8K
(input ratio 0.8). To evaluate the end-to-end effect of \methodname{}, we
select points on the existing Pareto line between 20 and 200 TPS/user as the
baseline. We then evaluate whether \methodname{} can push these points to a
better region, namely, higher output TPS/GPU at similar TPS/user, while
reporting the corresponding TTFT tradeoff separately. For
comparison, we keep the generation-server configuration fixed and apply
\methodname{} only to the context server, searching for improved points
primarily by varying the number of context GPUs. Unless otherwise stated, all
end-to-end results reported in this subsection do not yet include the
performance gain from the contention-mitigation optimization described in
Section~\ref{sec:design-optimizations}.

\begin{figure}[t]
  \centering
  \includegraphics[width=0.85\linewidth]{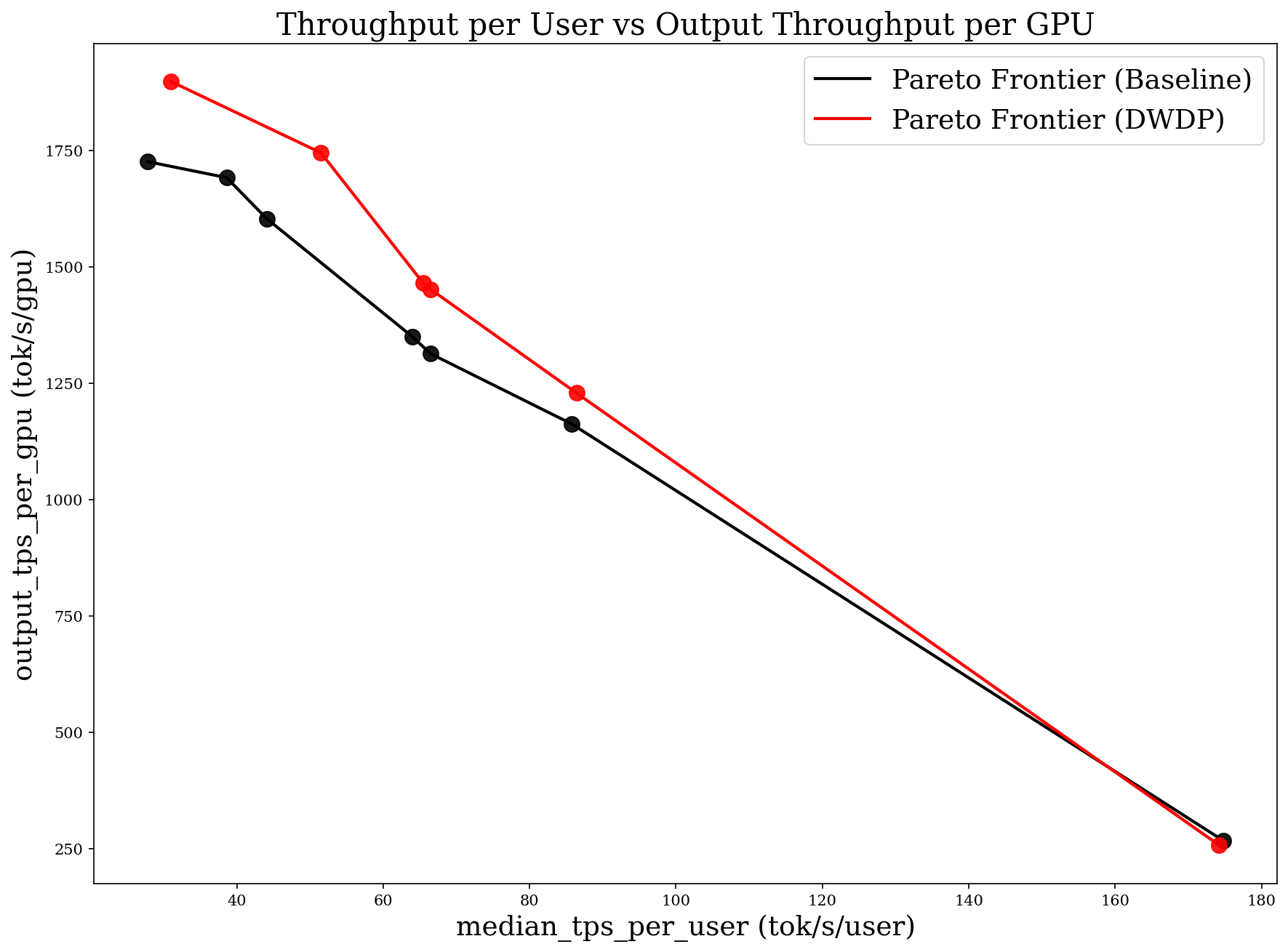}
  \caption{End-to-end Pareto frontier comparison between baseline and
  \methodname{}.}
  \label{fig:e2e-pareto-frontier}
\end{figure}

\paragraph{Analysis of TPS/User and Output TPS/GPU}
Figure~\ref{fig:e2e-pareto-frontier} shows that \methodname{} indeed pushes
the end-to-end Pareto points to better positions: at similar TPS/user, it
achieves higher output TPS/GPU than baseline across most of the target range. This
improvement in output TPS/GPU at comparable TPS/user is the primary end-to-end
benefit targeted by \methodname{}.
Table~\ref{tab:e2e-throughput-summary} summarizes the average speedup in each
TPS/user range. The gain is most pronounced at low TPS/user range. For
example, in the 20--30 TPS/user range, \methodname{} achieves 1.15$\times$
TPS/user speedup and 1.10$\times$ TPS/GPU speedup.

Comparing pareto points with similar TPS/user, we find that \methodname{}
typically uses fewer context GPUs than the baseline. This suggests that the
gain primarily comes from reduced context GPU demand. Based on prior knowledge, most of the points on the
baseline frontier are generation-bottlenecked, so accelerating only the
context phase does not directly translate into a large end-to-end throughput
increase. Instead, \methodname{} improves throughput per context GPU,
allowing the system to sustain comparable performance with fewer context GPUs
and therefore higher output TPS/GPU.
In some ranges, TPS/user also increases because using fewer context GPUs reduces
the aggregate request rate from the context phase, which in turn lowers the
runtime generation batch and improves per-user throughput at the generation
stage.

The serving-efficiency benefit becomes smaller at high TPS/user. In this region, the system is
more heavily generation-bottlenecked, and the context stage cannot accumulate
enough tokens to amortize \methodname{}'s prefetch overhead. As a result,
\methodname{} may provide limited efficiency benefit or even a slight regression.

\begin{table}[t]
  \centering
  \caption{End-to-end performance summary of \methodname{} across target TPS/user
  ranges.}
  \label{tab:e2e-throughput-summary}
  \begin{tabular}{lcc}
    \toprule
    TPS/user Range & Avg.\ \methodname{} TPS/user Speedup & Avg.\ \methodname{} TPS/GPU Speedup \\
    \midrule
    20--30 & 1.15 & 1.10 \\
    40--50 & 1.16 & 1.08 \\
    60--70 & 1.00 & 1.10 \\
    80--90 & 1.00 & 1.06 \\
    170--180 & 1.00 & 0.97 \\
    \bottomrule
  \end{tabular}
\end{table}

\begin{table}[t]
  \centering
  \caption{Median TTFT comparison across target TPS/user ranges.}
  \label{tab:e2e-ttft-summary}
  \begin{tabular}{lccc}
    \toprule
    TPS/user Range & TPS/GPU Speedup & Baseline TTFT (ms) & \methodname{} TTFT (ms) \\
    \midrule
    20--30 & 1.10 & 2538 & 8314 \\
    40--50 & 1.08 & 1919 & 7012 \\
    60--70 & 1.12 & 965 & 1640 \\
    80--90 & 1.06 & 1669 & 2280 \\
    170--180 & 0.97 & 494 & 660 \\
    \bottomrule
  \end{tabular}
\end{table}

\paragraph{Analysis of Time to First Token}
We also evaluate median TTFT, including queueing time, after applying
\methodname{}. For each baseline point, we select the \methodname{} point with
the closest TPS/user and compare their TPS/GPU and TTFT; the results are
summarized in Table~\ref{tab:e2e-ttft-summary}. Compared with the baseline,
\methodname{} increases TTFT across the evaluated TPS/user ranges. At high TPS/user, the system is
already heavily generation-bottlenecked, so the context stage cannot
accumulate enough tokens to amortize \methodname{}'s prefetch overhead. At
low TPS/user, TTFT can increase substantially for pairs with more aggressive
reductions in context GPU count. These regressions do not indicate lower
per-GPU context efficiency. Rather, reducing the context-side deployment can
lower the aggregate service rate of the context stage and worsen rate matching
between the context and generation stages. This in turn increases queueing
delay before the first token. We expect this issue can be mitigated by better request
matching in future work, especially because \methodname{} enables finer-grained
context configurations.

\section{Conclusion}

In this report, we presented \methodname{}, an inference parallelism
strategy that enables fully asynchronous execution across ranks. DWDP is
attractive for two reasons. Its asynchronous execution reduces synchronization
cost as modern LLM serving workloads become increasingly diverse and
imbalanced. It also supports more flexible GPU-count configurations, enabling
finer-grained resource provisioning and better rate matching between context
and generation servers in disaggregated serving.

A key requirement for DWDP is high-bandwidth communication. Systems such as
NVL72, which provide high-bandwidth communication between any pair of GPUs
within the domain, make it possible to hide remote weight prefetch behind
computation. At the same time, DWDP introduces new design challenges. An
efficient DWDP implementation requires co-design across communication, kernels, and
runtime. We will continue to improve the efficiency of DWDP in future work.

\bibliographystyle{unsrtnat}
\bibliography{references}

\clearpage
\appendix
\fancyhead[C]{\small\scshape Appendix}
\section{In-Depth Analysis of Communication-Computation Interference}
\label{sec:appendix-interference}

In \methodname{}, remote-weight prefetch is overlapped with computation. The overlap introduces hardware contention and reduces the expected gain.

\begin{figure}[H]
  \centering
  \includegraphics[width=0.75\linewidth]{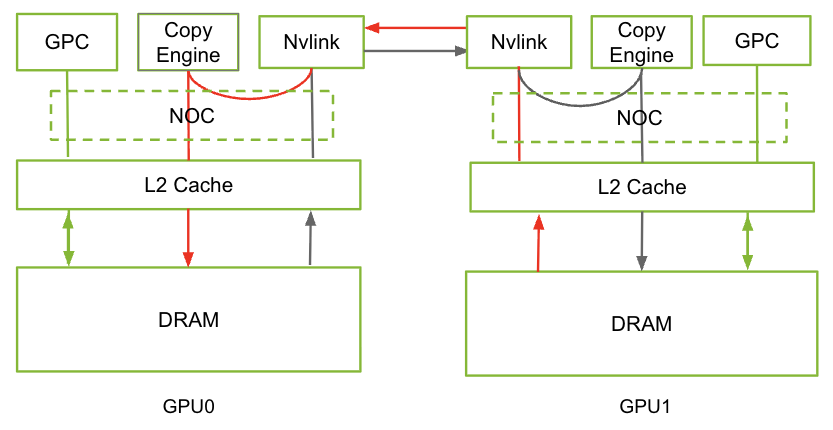}
  \caption{Copy-engine-driven remote weight pull across two GPUs. The transfer crosses NVLink and uses the NoC, L2 cache, and DRAM on both GPUs. Interference arises from shared NoC routing, L2 resources, DRAM bandwidth, and power budget.}
  \label{fig:ce-sm-interference}
\end{figure}

Figure~\ref{fig:ce-sm-interference} shows a copy engine (CE) on the destination graphics processing unit (GPU) pulling weights from a peer GPU over NVLink. The CE is a dedicated data-movement engine, so it does not occupy Streaming Multiprocessor (SM)-based computation resources. However, the transfer still traverses the Network-on-Chip (NoC), Level-2 (L2) cache, and dynamic random-access memory (DRAM) on both GPUs, while local SM kernels issue concurrent memory requests. As a result, remote-weight prefetch and computation can interfere at multiple levels of the hierarchy.

On the source GPU, the requested weight is fetched from DRAM, passes through L2 and the NoC, and is injected into NVLink; on the destination GPU, the data is received and written into its memory hierarchy. This path can interfere with computation through NoC arbitration, L2 pressure, DRAM bandwidth contention, and power contention. The last effect is especially important on modern high-power GPUs, which use dynamic voltage and frequency scaling (DVFS) to adjust operating frequency based on total graphics power (TGP). As a result, overlapping communication with heavy SM execution can trigger frequency throttling even before memory bandwidth is fully saturated.

Depending on the compute workload, the interference can be broadly divided into two types: contention with memory-bound kernels and contention with compute-intensive kernels.

\subsection{Contention with Memory-Bound Kernels: Memory Bandwidth Bottleneck}
We next analyze the effect of overlapping communication with memory-bound kernels. In the context stage, which often processes a large number of tokens, typical memory-bound workloads include quantization, copy, and element-wise vector kernels. In our DEP case, these kernels account for 18.2\% of the context-stage overhead. On Blackwell, the peak High-Bandwidth Memory (HBM) bandwidth is about 8\,TB/s, while NVLink~5 can consume up to 1800\,GB/s of aggregate read/write bandwidth. This corresponds to $1.8 / 8 = 22.5\%$ of the total memory bandwidth. Therefore, from the DRAM perspective, a memory-bound kernel can experience up to 22.5\% slowdown in the worst case when NVLink traffic is fully utilized. This upper bound is consistent with the end-to-end breakdown in Table~\ref{tab:context-breakdown}: the Others category, which is dominated by such memory-bound kernels, increases from $241.69$\,$\mu$s in DEP4 to $284.32$\,$\mu$s in DWDP4, corresponding to a slowdown of about $17.6\%$. The observed slowdown is smaller than the 22.5\% worst-case estimate for two reasons. First, communication does not overlap with all computation kernels. Second, L2 cache can absorb part of the activation traffic. More generally, both the copy engines and the SMs compete for access to HBM and the shared L2 cache, which can degrade both kernel execution and data transfer performance.

\subsection{Contention with Compute-Intensive Kernels: Power-Induced Frequency Bottleneck}
For highly compute-intensive workloads, such as the context-phase attention module, the dominant source of interference is not memory-system contention but power-induced frequency throttling.
As shown in Table~\ref{tab:context-breakdown}, the Attention kernel of DWDP4 is about $1.19\times$ slower than DEP4 for this module. This suggests severe interference when overlapping CE communication with heavy SM execution.

\paragraph{Bottleneck Analysis: Power and Frequency Contention.}
When heavy computation and communication kernels overlap, the total power draw can easily exceed the GPU's Thermal Design Power (TDP) limit. The attention module alone consumes about $96.7\%$ of the TDP limit, while two-sided communication draws about $30.5\%$ of the TDP limit, including an idle baseline of about $12.9\%$ of the TDP limit. The estimated total power during overlap is therefore roughly $96.7\% + 30.5\% - 12.9\% = 114.4\%$ of the TDP limit, which exceeds the power limit. This power capping triggers DVFS and leads to a significant reduction in GPU frequency.

\begin{figure}[H]
  \centering
  \includegraphics[width=\linewidth]{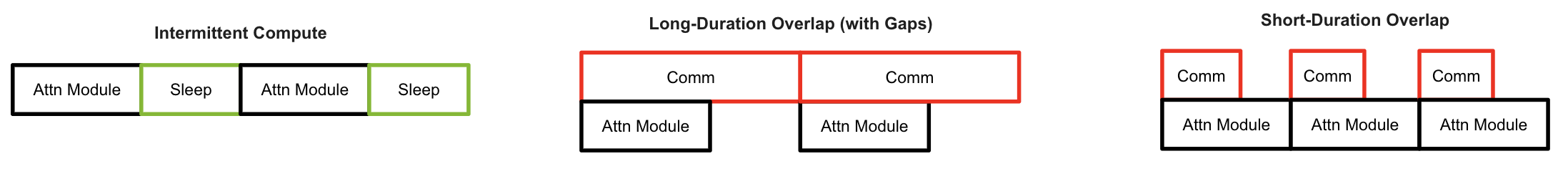}
  \caption{Illustration of the three communication patterns used in our overlap study. Intermittent Compute inserts large idle gaps between DeepSeek R1 attention modules to maximize power headroom, Long-Duration Overlap uses the longest communication windows while still preserving gaps between compute modules, and Short-Duration Overlap mimics the real DWDP workload with tightly scheduled compute and smaller communication tasks.}
  \label{fig:communication-patterns}
\end{figure}

\paragraph{Evaluation of Overlap Strategies.}
To understand the impact of this contention, we evaluate three scheduling configurations that differ only in how communication is arranged relative to the attention module. Figure~\ref{fig:communication-patterns} illustrates these communication patterns and highlights the difference between no overlap, coarse-grained overlap, and fine-grained overlap:
\begin{enumerate}
    \item \textbf{Intermittent Compute}: Large sleep gaps are inserted between DeepSeek R1 attention modules under the 16K-context setting, ensuring that each module runs with the best possible power headroom and without communication overlap.
    \item \textbf{Long-Duration Overlap (with Gaps)}: Each attention module overlaps with a long-duration CE communication task, yielding the longest overlap among the three patterns while still preserving large gaps between neighboring compute modules.
    \item \textbf{Short-Duration Overlap}: This pattern is closest to the real DWDP workload, where tightly scheduled attention modules overlap with smaller communication tasks.
\end{enumerate}

In Intermittent Compute, the sleep interval provides the DeepSeek R1 attention module with the largest power headroom and serves as a no-overlap baseline. Long-Duration Overlap maximizes the overlap duration, but the large gaps between neighboring compute modules still leave room for partial power recovery. Short-Duration Overlap is the most representative of the real DWDP execution pattern: compute modules are tightly scheduled, communication is broken into smaller transfers, and contention is repeatedly injected into an already power-constrained execution window.

\begin{table}[H]
  \centering
  \caption{GPU metrics for the DeepSeek R1 attention module under three communication-overlap patterns. Kernel time and GPU frequency are normalized to the Intermittent Compute baseline.}
  \label{tab:interference-metrics}
  \resizebox{\linewidth}{!}{
  \begin{tabular}{llccc}
    \toprule
    Category & Metric & Intermittent Compute & Long-Duration Overlap & Short-Duration Overlap \\
    \midrule
    \multirow{2}{*}{Power \& Thermal} & Normalized Kernel Time & 1.000 & 1.049 & 1.226 \\
     & Normalized GPU Frequency & 1.000 & 0.963 & 0.798 \\
    \midrule
    Memory & max Utilization & 50\% & 45\% & 47\% \\
    \midrule
    Memory & Derived Memory Access Count (GB) & 1.7 & 8.0 & 6.2 \\
    \midrule
    NVLink & Derived NVLink Memory Footprint (GB) & -- & 6.2 & 4.7 \\
    \bottomrule
  \end{tabular}}
\end{table}

\paragraph{Key Observations.}
Table~\ref{tab:interference-metrics} summarizes the GPU metrics and reveals four key properties of the interference behavior:

\begin{itemize}[leftmargin=1.5em]
    \item \textbf{Performance is tightly correlated with GPU frequency.} As shown in Table~\ref{tab:interference-metrics}, the normalized attention-kernel runtime increases to $1.049\times$ and $1.226\times$ under Long-Duration Overlap and Short-Duration Overlap, while the normalized GPU frequency drops to $0.963\times$ and $0.798\times$, respectively. Figure~\ref{fig:clock-perf-correlation} shows the same trend, indicating that attention-kernel time tracks GPU frequency across all three patterns.

    \begin{figure}[H]
      \centering
      \includegraphics[width=0.6\linewidth]{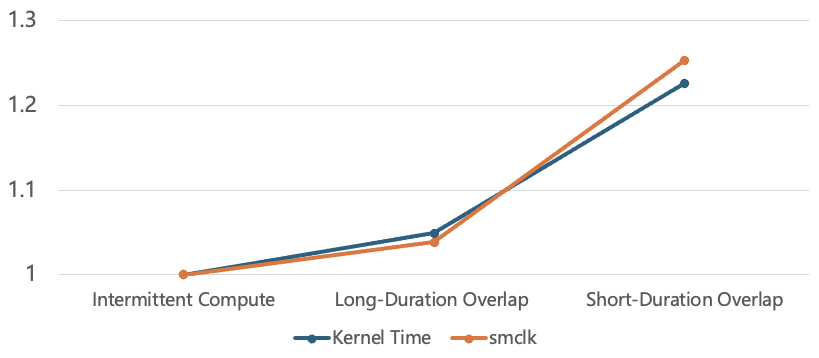}
      \caption{Normalized attention-kernel runtime and GPU frequency across the three overlap patterns. The two curves closely follow each other.}
      \label{fig:clock-perf-correlation}
    \end{figure}

    \item \textbf{L2 and DRAM bandwidth are not the primary bottlenecks.} In Table~\ref{tab:interference-metrics}, the maximum of memory utilization stays in a narrow range from $45\%$ to $50\%$ across all three patterns, indicating that neither resource approaches full saturation during overlap.

    \item \textbf{Overlapping CE communication introduces little L2 miss penalty.} If CE traffic significantly reduced the attention kernel's L2 hit rate, it would force more traffic between L2 and DRAM. Using the absolute counters in Table~\ref{tab:interference-metrics}, we subtract the NVLink communication footprint from the total derived memory-access count and obtain approximately $1.7$\,GB, $1.8$\,GB, and $1.6$\,GB for the three patterns. These values remain closely aligned, while the total L2 access volume is about $24$\,GB, indicating that overlapping communication causes little additional L2 miss penalty.

    \item \textbf{NVLink traffic is not the dominant source of slowdown.} Using publicly available profiling tools such as Nsight Systems and Nsight Compute, we monitored NVLink utilization and found that the instantaneous NVLink throughput does not show noticeable fluctuation when communication overlaps with the attention kernel. In other words, overlapping with the attention kernel does not significantly perturb the NVLink transfer rate; instead, the effective NVLink bandwidth is still primarily determined by system frequency. This observation suggests that NVLink or NoC contention is not the dominant cause of the slowdown.
\end{itemize}

In summary, when heavy compute overlaps with communication, the evidence points to \textbf{power-induced frequency throttling} rather than L2, DRAM, or NVLink contention. This suggests avoiding fine-grained overlap patterns that amplify power spikes.

\end{document}